\documentstyle[12pt]{article}
\newcommand{\be}{\begin{equation}}
\newcommand{\ee}{\end{equation}}
\newcommand{\ben}{\begin{eqnarray}\displaystyle}
\newcommand{\een}{\end{eqnarray}}
\newcommand{\bea}{\begin{eqnarray}\displaystyle}
\newcommand{\eea}{\end{eqnarray}}
\newcommand{\wh}{\widehat}
\newcommand{\al}{{( a )}}

\newcommand{\bet}{{( b )}}
\newcommand{\ten}{{(10)}}

\newcommand{\p}{\partial}

\begin{document}

\makeatletter
\renewcommand{\theequation}{\thesection.\arabic{equation}}
\@addtoreset{equation}{section}
\makeatother

\begin{titlepage}

\vfill

\begin{center}
   \baselineskip=16pt
   {\Large\bf  Five-dimensional heterotic black holes\\ and its dual IR-CFT}
   \vskip 2cm
    Hossein Yavartanoo
       \vskip .6cm
             \begin{small}
      \textit{Department of Physics, Kyung Hee University, Seoul 130-701, Korea}\\
        \textit{yavar@khu.ac.kr}
        \end{small}\\*[.6cm]
\end{center}

\vfill
\begin{center}
\textbf{Abstract}\end{center}
We analyze the possible dynamical emergence of IR conformal field theory describing the low-energy excitations of near-extremal black holes in five-dimensional compactification of heterotic strings. We find that, by tuning the mass and charges in such a way that the extremal black holes have a classically vanishing horizon area, the near-horizon develops an AdS$_3$ throat  and when we combine the low-energy limit with vanishing Newton coupling constant, the system has a dual conformal field theory  description.  We compare our results  with  c-extremization and the Kerr/CFT  predictions.

\begin{quote}
\end{quote}
\vfill
\end{titlepage}
\section{Introduction}

It is known that black holes are thermodynamical systems equipped with  temperature and entropy.  Therefore the main challenge  that a quantum theory of gravity should address is understanding what degrees of freedom count for the black hole's entropy.

Since string theory gives a framework for studying quantum gravity, we shall carry out our investigation in string theory. Recall that for a class of supersymmetric black holes in five-dimensional compactification of type IIB string theory, Strominger and Vafa computed the statistical entropy and found a precise agreement with the BekensteinÐHawking entropy \cite{Strominger:1996sh}.  This analysis has been generalized to several other extremal black holes and even non-extremal black holes in different theories of gravity in different dimensions.

For all extremal black holes that their near-horizon geometry contain a factor of AdS$_3$ or a quotient thereof, a simpler microscopic model is available. Since quantum gravity in asymptotically AdS$_3$ can be described by a two-dimensional conformal field theory (2d-CFT),  the entropy and the Hawking radiation of these extremal or near-extremal black holes can be accounted using only the universal properties of a dual 2d-CFT which is defined in the near-horizon region(for reviews, see \cite{Simon:2011zza}). 
By embedding these black hole solution to string/M-theory, ultraviolet completions of these AdS/CFT correspondences can be constructed using string theory.

Considerable progress has been made in the last few years in reproducing the entropy of the rotating black holes using dual field theory description. For these black hole solutions the dual field theory has many properties in common with 2d-CFT                                                             (see \cite{Compere:2012jk}  for a recent reveiw). The idea underlying this so-called Kerr/CFT correspondence, is generalized to a larger class of black holes in different theories of gravity in four and higher dimensions. 

The only essential requirement for these extensions of the Kerr/CFT correspondence is the the presence of a U(1) axial symmetry associated with angular momentum.

Despite the success in computing the BekensteinÐHawking entropy of extremal black holes, either using the Kerr/CFT correspondence or the entropy function formalism based on the enhancement of symmetry of their near-horizons (e.g. see \cite{senpapers, senpapers2,Sen-AdS2/CFT1}), there are arguments suggesting these AdS$_2$ geometries do not generically represent a decoupled conformal field theory. Even if they would, the AdS/CFT correspondence would suggest these theories may be dynamically trivial \cite{Horowitz-Marolf-Harvey}, in the sense that they only contain degeneracy of the vacuum in their spectrum. So the Kerr/CFT conjecture, in its current status, is more a suggestion for a possible pair of theories dual to each other and many things should be understood better to establish the proposal as a concrete duality.

The goal of this note is studying the circumstances under which the near-horizon limit of extremal black holes exhibit non-trivial dynamics and analyzing
the emergence of low-energy dynamical conformal symmetry.  In this paper we consider certain near-extremal static charged black holes in five-dimensional compactification of heterotic string and analyze its near-horizon geometry to study the low-energy excitations.

Motivated by results  in \cite{SheikhJabbaria:2011gc},  we focus on the extremal black hole solution which has a vanishing horizon area. They belong to the family of Extremal Vanishing Horizon (EVH) black holes whose near-horizon geometry develops a local AdS$_3$  throat.  This suggests the possible existence of an (IR) dual 2d-CFT    which captures the low-energy dynamics around the background of the EVH black hole.

 In this work we first review dimensional reduction of ten-dimensional heterotic string  theory. We adopt the notation and terminology from \cite{Sen:1994wr,Sen:1994fa}. 
 In Section 3 we review the three charge static black hole solution in the five-dimensional theory.  The black hole solution is reviewed in   \cite{Youm:1997hw}.  
 In Section 4 we study near-horizon geometry of the five-dimensional black hole solution and its uplifting to ten dimensions.  We use the Kerr/CFT correspondence to explore the chiral CFT dual to these geometries.  In Section 5 we analyze the EVH limit of these black hole solutions. We show that near-horizon geometry of the EVH black holes contain a pinching AdS$_3$ factor which in near-horizon geometry of the near-EVH black holes this pinching AdS$_3$ throat turns into a pinching BTZ black hole. This pinching BTZ has parametrically the same entropy as the original near-EVH black hole. 
 In Section 6 we discuss possible (IR) dual CFT and compare the results with Kerr/CFT predictions in Section 4. We devote Section 7 to a discussion.

\section{Dimensional Reduction of the Ten-Dimensional Theory}

In this section we review some basic properties of five-dimensional effective field theory, resulting from compactification of ten-dimensional  heterotic string theory  on a five
dimensional torus $T^4\times S^1$.    Low-energy effective action of heterotic string theory in ten dimensions is an ${\mathcal N} = 1$
supergravity theory with 16 real supercharges, coupled to ${\mathcal N}=1$ super Yang-Mills theory in ten
dimensions. The bosonic sector of the theory contains  a dilaton $\Phi^{(10)}$, a metric tensor $G^{(10)}_{MN}$, a 2-form gauge field $B^{(10)}_{MN}$ and SO(32) or E$_8\times$E$_8$  Yang-Mills gauge fields $A_M^{(10)}$.

At a generic point in the moduli space, only the abelian
gauge fields give rise to massless fields in five dimensions, therefore it
is enough to restrict ourselves to the U(1)$^{16}$ part of the gauge group of the ten-dimensional theory.  The bosonic part of the ten-dimensional action is given by

\ben \label{new1}
S={1\over 2\kappa_{10}^2} \int d^{10}x \sqrt{ - G^{(10)}}\,
e^{-\Phi^{(10)}}\Big(R^{(10)} +G^{\ten MN}\p_M\Phi^\ten \p_N\Phi^\ten
- {1\over 12} H^{(10)}_{MNP}
H^{(10)MNP}
\nonumber \\
 - {1\over 4} F^{(10)I}_{MN} F^{(10)IMN}\Big),\quad  0 \leq M,N,P \leq 9,\;\; 1\leq I \leq 16,
\een
where

\ben \label{new2}
F^{(10)I}_{MN} &=& \p_M A^{\ten I}_N - \p_N A^{\ten I}_M \nonumber\\
H^{(10)}_{MNP} &=& (\p_M B^\ten_{NP} -{1\over 2} A_M^{\ten I}
F^{\ten I}_{NP}) + \hbox{cyclic permutations in $M$, $N$, $P$}. 
\een
Now let us rewrite, for  later convenience,  the action in the Einstein frame, in which the Ricci scalar is not multiplied by the scalar field.  This can be achieved by performing the following conformal transformation:
\be
G^{(10)}_{MN} = e^{\frac{1}{4}\Phi^{(10)}} G^{(10E)}_{MN}
\ee
and one gets the action in the Einstein frame to be
\bea
\label{action10dE}
S={1\over 2\kappa_{10}^2} \int d^{10}x \sqrt{ - G^{(10E)}}\,
\Big(R^{(10E)} -\frac{1}{8}G^{(10E) MN}\p_M\Phi^\ten \p_N\Phi^\ten
\nonumber \\
- {1\over 12} e^{-\frac{1}{2}\Phi^{(10)}} H^{(10)}_{MNP}
H^{(10)MNP} - {1\over 4} e^{-\frac{1}{4}\Phi^{(10)}} F^{(10)I}_{MN} F^{(10)IMN}\Big),
\eea

Let us analyze briefly massless field content in bosonic sector after a Kaluza-Klein compactification to five dimensions.  We use Greek incises $\mu, \nu, \cdots$ to label the five-dimensional space-time while using Latin indices $m, n , \cdots$ to label the internal space directions. 
From 16 U(1) gauge fields A$_M^{(10) I}$ in ten dimensions, we get  16 U(1) gauge fields A$_\mu^{10+I}$ in five dimensions and 80  scalar fields $\hat{\mathrm A}_m^I$. 
From the 2-form B$_{MN}$ one obtains a 2-form B$_{\mu\nu}$ , five U(1) gauge fields A$_{\mu}^{m+5}$ and ten scalars  $\hat{\mathrm B}_{mn}$. 
The ten-dimensional metric  G$_{MN}^{(10)}$ gives a five-dimensional metric tensor G$_{\mu\nu}$, five U(1) gauge fields A$_{\mu}^m$ and 15 scalars $\hat{\mathrm G}_{mn}$.    In summary, we have a dilaton $\Phi$, a metric G$_{\mu\nu}$, an antisymmetric tensor field B$_{\mu\nu}$, 26 KK U(1) gauge fields, and 105  scalar fields.

In five dimensions,  the NS-NS 3-form field strength is dual to a 2-form field strength. This 2-form can be considered as the field strength of a new U(1) gauge field $A_{\mu}$. 
Therefore the black hole solution in this theory carries an additional charge which is associated with the 2-form field $B_{\mu\nu}$ as well as charges of the 26 U(1) gauge fields. 
Thus, the most general black hole in heterotic string on T$^5$ can be  parameterized by 27 charges, two angular momenta and its temperature.

It is convenient to introduce the
{\it five-dimensional fields}  $\wh G_{mn}$, $\wh B_{mn}$, $\wh
A^I_m$, $\Phi$, $A_\mu^\al$, $G_{\mu\nu}$ and $B_{\mu\nu}$ ($1\le
m\le 5$, $0\le \mu\le 4, 1\leq a \leq 26$) 
with the following relations:

\ben\label{1.2}
&& \wh G_{mn}  = G^\ten_{m+4,n+4}, \quad  \wh B_{mn}  =
B^\ten_{m+4, n+4}, \quad  \wh A^I_m  = A^{\ten I}_{m+4},
\nonumber \\
&& A^{(m)}_\mu  = \wh G^{mn} G^\ten_{n+4,\mu}, \quad
A^{(I+10)}_\mu = -(A^{\ten I}_\mu - \wh A^I_n
A^{(n)}_\mu), \nonumber \\
&&  A^{(m+5)}_\mu = 
B^\ten_{(m+4)\mu} - \wh B_{mn} A^{(n)}_\mu + \wh A^I_m
A^{(I+12)}_\mu, \nonumber \\
&& G_{\mu\nu} = G^\ten_{\mu\nu} - G^\ten_{(m+4)\mu} G^\ten_{(n+4)\nu} \wh
G^{mn}, \nonumber \\
&& B_{\mu\nu} = B^\ten_{\mu\nu} - \wh B_{mn} A^{(m)}_\mu
A^{(n)}_\nu - {1\over 2} (A^{(m)}_\mu A^{(m+5)}_\nu - A^{(m)}_\nu A^{(m+5)}_\mu),
\nonumber \\
&& \Phi = \Phi^\ten - {1\over 2} \ln\det \wh G, \quad \quad
\quad 1\le m, n \le 5, \quad
0\le \mu, \nu \le 4, \quad 1\le I \le 16. 
\een

Here $\wh G^{mn}$  is the inverse of the matrix $\wh G_{mn}$.
One can combine the scalar fields $\wh G_{mn}$, $\wh B_{mn}$, and
$\wh A_m^I$ into an  $O(5,21)$ matrix valued scalar field $M$.
To do this, we define a $5\times 5$ matrix $\wh C_{mn} = {1\over 2} \wh A^I_m
\wh A^I_n$  and regard $\wh G_{mn}$, $\wh B_{mn}$ and $\wh A^I_m$ as
$5\times 5$, $5\times 5$, and $5\times 16$ matrices
respectively.   Then we define $M$ to be following
$26\times 26$-dimensional matrix

\bea
M = \pmatrix{\displaystyle  \wh G^{-1} & \wh G^{-1} (\wh B + \wh
C) & \wh G^{-1}\wh A   \cr (-\wh B + \wh C) \wh G^{-1}  & (\wh G
- \wh B +
\wh C) \wh G^{-1} (\wh G + \wh B + \wh C)  & (\wh G -\wh B +\wh
C)  \wh G^{-1} \wh A    \cr  \wh A^T \wh G^{-1} &  \wh A \wh G^{-1} 
(\wh G + \wh B +\wh
C) &  I_{16} + \wh A^T \wh G^{-1} \wh A\cr  } . \eea
satisfying

\bea
M L M^T = L, \quad \quad  M^T=M,  \quad \quad L =\pmatrix{0 & I_5
& 0  \cr I_5 & 0 & 0 \cr 0 & 0 & -I_{16}},  
\eea
where  $I_n$ denotes the identity matrix of rank $n$.

To get the effective action which governs the dynamics of the massless
fields in the five dimensions, we
substitute the expressions for the ten-dimensional fields in
terms of the five-dimensional fields in action (\ref{new1}), then taking all
field in the action to be independent of the internal directions. 

The
result is
\ben 
S &=& \frac{1}{16 \pi G_5} \int d^5 x \sqrt{-  G} \, e^{-\Phi} \big[ R +
G^{\mu\nu}
\p_\mu \Phi \p_\nu\Phi -{1\over 2 . 3!} G^{\mu\mu'} G^{\nu\nu'}
G^{\rho\rho'} H_{\mu\nu\rho} H_{\mu'\nu'\rho'} \nonumber \\
&&\quad\quad  - \frac{1}{2.2!}G^{\mu\mu'} G^{\nu\nu'} F^\al_{\mu\nu} (LML)_{ab}
F^\bet_{\mu'\nu'} + {1\over 8} G^{\mu\nu} Tr (\p_\mu M L \p_\nu
M L) \big] 
\een
where G$_N$ = G$_N^{(10)}/$ V$_5$ is the effective 5-dimensional Newton constant  and V$_5$ is volume of T$^5$. 
Field strengths $F^{(a)}$ and $H$ are given by
\ben 
\label{fields}
&& F^\al_{\mu\nu} = \p_\mu A^\al_\nu - \p_\nu A^\al_\mu \nonumber \\
&&H_{\mu\nu\rho} = (\p_\mu B_{\nu\rho} + {1\over 2} A^\al_\mu
L_{ a  b } F^\bet_{\nu\rho}) + \hbox{cyclic permutations of
$\mu$, $\nu$, $\rho$}.  
\een
From now on we take
$\int d^5 x=1$, where $x^m$ ($5\le m\le 9$) denotes the coordinates
labeling the five-dimensional torus. Therefore we get $G_5=G_N$.

To write down the five-dimensional action in the Einstein frame, we introduce
the canonical metric,
\bea 
g_{\mu\nu} = e^{-2\Phi \over 3} G_{\mu\nu}, 
\eea
and we also use the convention that all indices are lowered or raised  with respect to this canonical metric. The five-dimensional effective action in the  Einstein frame is given by
\ben 
\label{action2}
S &=& \frac{1}{6\pi G_5 } \int d^5 x \sqrt{-g} \,  \big[ R_g -\frac{1}{3}
g^{\mu\nu}
\p_\mu \Phi \p_\nu\Phi -{1\over 2 . 3!} e^{-\frac{4}{3}\Phi}g^{\mu\mu'} g^{\nu\nu'}
g^{\rho\rho'} H_{\mu\nu\rho} H_{\mu'\nu'\rho'} \nonumber \\
&&\quad\quad  - \frac{1}{2.2!} e^{-\frac{2}{3}\Phi}g^{\mu\mu'} g^{\nu\nu'} F^\al_{\mu\nu} (LML)_{ab}
F^\bet_{\mu'\nu'} + {1\over 8} g^{\mu\nu} Tr (\p_\mu M L \p_\nu
M L) \big].
\een

\section{The black hole solution}

Let us consider a compactification on T$^4 \times$ S$^1$  in which T$^4$ is completely factorized, i.e all KK gauge fields which are obtained from T$^4$-indices are taken to be zero. 
Therefore in bosonic sector of the theory, non-vanishing dynamical massless fields are dilaton $\Phi$, metric $g_{\mu\nu}$, 2-form B$_{\mu\nu}$, modulus $\hat{G}_{55}$, and two U(1) gauge fields A$^{(i)}_{\mu}$, i = 1,6.

A class of extremal black hole solutions in this theory have been constructed in \cite{Cvetic:1998xh}\footnote{This black hole has been constructed in \cite{Cvetic:1996xz} first, but the solution presented there is incorrect.}  In five dimensions, the black hole carries only electric charges of U(1) gauge fields and the 3-form field strength $H_{\mu\nu\alpha}$ which is the Hodge-dual to a 2-form field strength $F$ in the following way:
\bea 
H^{\mu\nu\alpha}= -\frac{e^\frac{4\Phi}{3}}{2\sqrt{-g}}\epsilon^{\mu\nu\alpha\beta\rho} F_{\beta\rho} 
\eea
where $F$ itself is the field strength of a new U(1) gauge field $A$ and the black hole solution carries the charge associated with the 2-form field $B$ as well.  In summary we have the following bosonic fields in our five-dimensional theory
\bea   \nonumber
&& \hspace{-9mm} \Phi=\Phi^{(10)}-\frac{1}{2}\ln G_{55}^{(10)}, \quad \;\;\;  G_{\mu\nu}=G_{\mu\nu}^{(10)}-\left(G_{55}^{(10)}\right)^{-1}G_{5\mu}^{(10)} G_{5\nu}^{(10)},\\  \nonumber
&& \hspace{-9mm} S=e^{-\Phi}, \quad T=\sqrt{G_{55}^{(10)}}, \quad A_{\mu}^{(1)} = \left(G_{55}^{(10)}\right)^{-1}G_{5\mu}^{(10)} , \quad A_{\mu}^{(6)} = B_{5\mu}^{(10)},\\  \nonumber
&& \hspace{-9mm} B_{\mu\nu} =  B^{(10)}_{\mu\nu}  -{1\over 2} \left(A_{\mu}^{(1)}A_{\nu}^{(6)}  - A_{\nu}^{(1)}A_{\mu}^{(6)}  \right) 
\eea
Using the action (\ref{action2}), we obtain the following effective action which governs the dynamics of above bosonic fields:
\bea
\label{action3}
&& S= \frac{1}{16\pi G_N}\int d^5x \sqrt{-g}\;  \bigg[ R_g-\frac{1}{3} S^{-2}(\partial S)^2-T^{-2} (\partial T)^2
\cr\cr && \hspace{15mm}-\frac{1}{4}S^{2\over 3}T^2 (F^{(1)})^2- \frac{1}{4}S^{2\over 3}T^{-2} (F^{(6)})^2-\frac{1}{12}S^{4\over 3} H^2\bigg],
\eea
where from (\ref{fields}) the field strengths $H$, $F^1$ and $F^6$ are defined by
\bea 
&&H_{\mu\nu\rho} = \p_\mu B_{\nu\rho} + {1\over 2} \left(A^{(1)}_\mu  F^{(6)}_{\nu\rho} -A^{(6)}_\mu  F^{(1)}_{\nu\rho}  \right) + \hbox{cyclic permutations of
$\mu$, $\nu$, $\rho$},  \nonumber \\
&& F^{(1)}_{\mu\nu} = \p_\mu A^{(1)}_\nu - \p_\nu A^{(1)}_\mu ,\quad  F^{(6)}_{\mu\nu} = \p_\mu A^{(6)}_\nu - \p_\nu A^{(6)}_\mu 
\eea

We note that the truncated action (\ref{action3}) has reduced supersymmetry of the original theory from ${\mathcal N }= 4$ to ${\mathcal N} = 2$. It also has a T-duality group which is generated by action $T \rightarrow 1/T$  and $ A^{(1)} \leftrightarrow A^{(6)}$.

In terms of the non-trivial five-dimensional bosonic fields, the black hole solution is of the form
\bea
 \label{5dBHsolution}
&&T=\frac{(r^2+2m\sinh^2\delta_1)^{1\over 2}}{(r^2+2m\sinh^2\delta_2)^{1\over 2}}, \quad  S=\frac{(r^2+  2m\sinh^2\delta_1)^{1\over 2}(r^2+2 \sinh^2\delta_2)^{1\over 2}}{(r^2+2 \sinh^2\delta_3)},  \nonumber \\
&&A^{(1)}=\frac{m\sinh2\delta_1 dt}{r^2+2m\sinh^2\delta_1} ,\quad\;  A^{(6)}=\frac{m\sinh2\delta_2 dt}{r^2+2m\sinh^2\delta_2},  \quad \; B_{ \psi \varphi} =m\sinh2\delta_3 \cos^2\theta,\nonumber \\
&& ds^2=\Delta^{1\over 3}\bigg[-\frac{r^2(r^2-2m)}{\Delta}dt^2+ \frac{dr^2}{r^2-2m} +
d\Omega_3^2\bigg] 
\eea

where
\bea
d\Omega_3^2=d\theta^2+\sin^2\theta d\varphi^2 +\cos^2\theta d\psi^2,\quad  \Delta= \prod_i (r^2+2m\sinh \delta_i^2),\;\;\; i=1, 2, 3.
\eea
Note that the solution has the inner and outer horizons which are located at $r=0$ and $r=\sqrt{2m}$ respectively. The U(1) charges Q$_i$'s, the ADM mass M for above black hole solution are given by:
\bea
M=\frac{\pi m}{4G_N} \sum_i \cosh 2\delta_i ,\quad Q_i= m \sinh2\delta_i
\eea
where $Q_1$ and $Q_2$ are electric charges under gauge fields $A^{(1)}$ and $A^{(6)}$ and in 10 dimensions  they are momentum and winding number of the elementary string wound around $S^1$ circle.  While $Q_3$ is the magnetic charge under 2-form gauge field $B_{\mu\nu}$ and it is interpreted as charge of NS5 branes wound around T$^5$. For the general value of $m$, the black hole solution is non-extremal and breaks all supersymmetries of the theory. The entropy and Bekenstein-Hawking temperature are given by
\bea
\label{BHET}
S_{\mathrm BH}=\frac{\pi^2 \sqrt{2m^3}} {G_N}\prod_{i=1}^3 \cosh\delta_i ,\quad T_{\mathrm BH}=\frac{1}{2\pi \sqrt{2m}}\; \prod_{i=1}^3  {\mathrm sech}\;\delta_i.
\eea
The extremal limit is given by $m=0$  while keeping black hole charges finite which is corresponding to scale $\delta_i$ to infinity. In this limit the Bekenstein-Hawking temperature vanishes and for $\delta_1\delta_2 >0$ the corresponding solution is 1/4-BPS  \cite{Ferrara:1997ci}.  For extremal black hole solutions, it is more convenient to use following parameters: 
\bea
|q_1|\equiv \frac{m}{4G_N} \sinh^2\delta_1,\quad |q_2|\equiv \frac{m}{4G_N} \sinh^2\delta_2,\quad  |p| \equiv 8 \pi^2 m\sinh^2\delta_3.
\eea
which give the fundamental string charges, winding number and the number of NS5 branes respectively. 
Therefore for the extremal black hole  Bekenstein-Hawking entropy (\ref{BHET}) can be expressed as
\bea
S_{\mathrm BH}=2\pi \sqrt{|pq_1q_2|} 
\eea

Near-horizon behavior of the extremal black hole solution is given by:
\bea
&&\hspace{-5mm} ds^2=R_{\mathrm AdS_2}^2\left(-\rho^2 d\tau^2+\frac{d\rho^2}{\rho^2}+ 4 d\Omega_3^2   \right),  \quad F^{(1)}_{\tau\rho}=\frac{1}{4\pi q_1}\sqrt{|pq_1q_2|} , \quad  F^{(6)}_{\tau\rho}=\frac{1}{4\pi q_2}\sqrt{|pq_1q_2|},   \nonumber \\ \label{NHGads2}
&& \hspace{-5mm}  H_{\theta\varphi\psi}=\frac{p}{2\pi^2}\sin\theta\cos\theta, \;\;\; S=32 \pi^2 G_N \frac{\sqrt{|q_1q_2|}}{|p|} ,  \;\;\; T=\sqrt{|\frac{q_1}{q_2}|}, \quad R_{\mathrm AdS_2}= \left(\frac{G_N^{2}\;|pq_1q_2|    }{4\pi^{2}} \right)^{1\over 6}
\eea
 The near-horizon metric is AdS$_2 \times$S$^3$ and when all charges are finite and non-vanishing, R$_{\mathrm AdS_2}$ is finite. Indeed it is easy to check that for the extremal black hole solution, horizon which is located at $r = 0$, is a regular surface, with all curvature invariants being finite and well defined on it. We call such black hole solutions large extremal black holes. The near-horizon geometry (\ref{NHGads2}) is completely determined by charges $Q_i$, and is independent of asymptotic values of moduli. This is an example of the attractor mechanism. It is also easy to check that the near-horizon configuration (\ref{NHGads2}) itself is an exact solution of equations of motion. There is a supersymmetry enhancement on the near-horizon and geometry (\ref{NHGads2}) preserves all supersymmetries of the five-dimensional theory. 
 
For our later purpose, let us uplift the five-dimensional solution (\ref{5dBHsolution}) into ten-dimensions.  In the string frame non-zero bosonic fields are given by
\bea
\label{10ds}
&& ds^2=\frac{-r^2(r^2-2m)dt^2}{(r^2+2m\sinh^2\delta_1)(r^2+2m\sinh^2\delta_2)} +\frac{(r^2+2m\sinh^2\delta_3)dr^2}{r^2-2m} +(r^2+2m\sinh^2\delta_3) d\Omega_3^2\nonumber \\ 
&& \hspace{12mm}+\frac{r^2+2m\sinh^2\delta_1}{r^2+2m\sinh^2\delta_2}\left(dx_5+\frac{m\sinh 2\delta_1}{r^2+2m\sinh^2\delta_1}\;dt\right)^2 +\sum_{i=6}^9 dx_i^2, \nonumber \\ 
&&  e^{\Phi^{(10)}}=\frac{r^2+2m\sinh^2\delta_3}{r^2+2m\sinh^2\delta_2},\quad  B_{t5}^{(10)}=-\frac{m\sinh2\delta_2 }{r^2+2m\sinh^2\delta_2},\quad B_{\phi\psi}^{(10)}=-m\sinh2\delta_3\cos^2\theta.
\eea

To get the near-horizon geometry of the extremal limit of solution (\ref{10ds}), we set
\bea 
r=\frac{2G_N}{\pi }\sqrt{|pq_1q_2| \epsilon \rho } ,\quad   t =\frac{\tau}{\epsilon},\quad x_5=\tilde{x_5}-t, 
\eea
and take the limit $\epsilon \rightarrow 0$. The shift from  $x_5$ to $\tilde{x_5}$  makes vector $\partial_t$ tangent to the horizon. In
other words, the coordinates co-rotate with the horizon. The result is
\be
\label{NHads2}
\label{NHG10dS} ds^2=\frac{R^2}{4}\left[ -\rho^2 d\tau^2 +\frac{d\rho^2}{\rho^2}+\alpha^2 \left(d\tilde{x}_5-\frac{\rho }{\alpha} d\tau \right)^2\right] +R^2 d\Omega_3^2 + \sum_{i=6}^9 dx_i^2,
\ee
where 
\bea\
R^2 = \frac{|p|}{4\pi^2},\quad \alpha=\frac{4\pi q_1}{\sqrt{|pq_1q_2|}} 
\eea
In above geometry, the AdS$_2$ part along with circle $\tilde{x}_5$, which is fibered over the AdS$_2$ base, form a space-time which is asymptotically locally isometric to AdS$_3$ and has a {\it `null cylinder'} as its boundary. In fact, this geometry is the space-like self-dual AdS$_3$ orbifold of Coussaert and Henneaux \cite{Coussaert:1994tu,c-h}. This is the same geometry as which appears in near-horizon limit of the extremal BTZ black hole.  To see that, let us  consider an extremal BTZ black hole
\bea
ds^2=-\frac{(r^2-r_h^2)}{l_3^2r^2}dt^2+\frac{l_3^2r^2 dr^2}{(r^2-r_h^2)^2}+r^2\left(d\varphi-\frac{r_h^2}{r^2}\frac{dt}{l_3}\right)^2  
\eea
with
\bea
l_3=\frac{\sqrt{|pq_1q_2|}}{2q_1},\quad\quad r_h=2\sqrt{|\frac{q_1}{q_2}|},   
\eea
and taking the near-horizon limit
\bea
r^2=r_h^2+\epsilon\rho, \quad t=\frac{\tau}{\epsilon}, \quad \varphi= 2\pi \tilde{x}_5-\frac{\tau}{\epsilon l_3}, \quad \epsilon\rightarrow 0,    
\eea
we get the three dimensional part of metric (\ref{NHG10dS}), spanning  by $(\tau, \rho, \tilde{x}_5)$.  The appearance of the AdS$_3$ throat in the near-horizon geometry, indeed is very suggestive of having a 2d-CFT  dual description to physics on this geometry. However we should note that, what we obtain in the near-horizon is not a round AdS$_3$, but it is an orbifold of AdS$_3$. 

For the effective field theory on AdS$_3$, in compactification of heterotic string theory, one could find central charges c$_{R,L}$.  Then the statistical entropy would be simply given by the Cardy formula (for a review of related discussions see \cite{Prester:2010cw}). 

There are two methods which were used to find central charges. The first method is based on direct sigma model calculation \cite{Kutasov:1998zh}. In the second method, central charges are indirectly given by using anomaly inflow arguments \cite{Kraus:2005vz,Kraus:2006wn}. In the first method, which is historically earlier method, one treats heterotic string theory on the  backgrounds with AdS$_3$ factors. Then by relying on explicit realizations of (0,4) supersymmetry and using AdS/CFT correspondence we can obtain relevant central charges \cite{Kutasov:1998zh, Barnich:2006av}.  When $q_2 > 0$ the central charges are given by
\bea\label{clcr}
c_R=6|q_2 p|,\qquad  c_L=6|q_2|(|p|+3),
\eea
while for $q_2 < 0$ the only change is $c_R\leftrightarrow c_L$. Note that, to get the above result one needs to add Chern-Simons term to the effective action (\ref{new1}).  Using the Cardy formula we obtain the statistical entropy in the BPS case ($q_1q_2>0$), 
\bea
S_{CFT}=2\pi\sqrt{q_1q_2(|p|+3)},   
\eea
while in the non-BPS case ($q_1q_2<0$), it is given by 
\bea
S_{CFT}=2\pi\sqrt{|q_1q_2 p|} .
\eea
The result in the BPS case agrees with statistical entropy obtained by direct microstate counting \cite{Sen:2007qy,Castro:2008ys}.

As we have mentioned earlier, the near-horizon geometry (\ref{NHG10dS}), does not have a round AdS$_3$ factor. But since the self-dual orbifold  is asymptotically locally AdS$_3$, we expect the dual field theory to be a two dimensional conformal field theory, but defined on a boundary of the null cylinder.

In \cite{us} is has been argued that this geometry is dual to DLCQ of a 2d-CFT which is a chiral CFT.  Since the eigenvalues of left and right excitation operators are scaled with opposite boost factors, the DLCQ creates an infinite mass gap in one of the sectors, say the $\tilde{L}_0$ sector, if we intend to keep the mass scale of the
other sector finite.  Due to the infinite mass gap, we cannot excite the $\tilde{L}_0$  sector and it is to be set to its ground state. On the other hand,  the AdS$_2$ factor which cannot be excited and which is the geometric manifestation of the frozen $\tilde{L}_0$ sector. The excitations in the dynamical $L_0$ sector crucially involve the $\tilde{x}_5$ coordinate.

  \section{Extremal black hole/CFT description}
 It was argued that the entropy of rotating black holes can be obtained from the Cardy formula of a 2d-CFT, living in the boundary of near-horizon geometry of the black hole (for a recent review, see \cite{Compere:2012jk}). This has been generalized to the {\it extremal black hole/CFT} correspondence in \cite{Hartman:2008pb,Chow:2008dp}. 
 
 Consider a black hole solution with the following near-horizon field configuration
 \bea
 && ds^2=A\left( -\rho^2 dt^2 +\frac{d\rho^2}{\rho^2}\right) +\sum_{\alpha=1}^n F_{\alpha}dy_{\alpha}^2+\sum_{i, j=1}^{n-1+\epsilon} \tilde{g}_{ij} \tilde{e}_i \tilde{e}_j \nonumber \\
 &&  \tilde{e}_i=d\phi_i+k_i\rho dt, 
 \eea
where  functions $A$, $F_{\alpha}$ and  $\tilde{g}_{ij}$  depend only on the latitudinal coordinates. When a certain boundary condition is applied to this near-horizon geometry, one can find that the asymptotic symmetry group includes a Virasoro algebra extension for each of the compact $U(1)$ isometries.  Corresponding central charges are given by
\bea
c_i=\frac{6k_iS_{BH}}{\pi}, 
\eea
where $S_{BH}$ is the entropy of the original extremal black hole solution. This holds for all $i$'s \cite{Chow:2008dp}. Using the Cardy  formula,  we can evaluate statistical entropy 
\bea 
S_{stat.} = 2\pi\sqrt{\frac{c_i}{6}\left(L_0^i-\frac{c_i}{24}\right)} = \frac{\pi^2}{3}c_iT_i=S_{BH},\qquad {\mathrm for\;each\;} i 
\eea
where the CFT temperature is defined by
\bea
k_i= \frac{1}{2\pi T_i}, \quad T_{i}=-\frac{T'^0_{BH}}{\Omega'^0_{i}},\quad\quad T'^0_{BH}=\frac{\partial T_{BH}}{\partial r_+}|_{r_+=r_0},\quad  \Omega'^0_i=\frac{\partial \Omega_i}{\partial r_+}|_{r_+=r_0}, 
\eea
$T_H$ and $\Omega_i$ are  the Hawking temperature and angular velocities, respectively,  and $r_0$ is the extremal value for the outer horizon radius $r_+$.  Although in the Kerr/CFT correspondence there could be several chiral CFT's with various central charges associated with each U(1), however these chiral CFT's are expected to be related by string theory/supergravity dualities and there is essentially only one gauge invariant CFT.

The near-horizon geometry (\ref{NHads2})  fits the above discussion.  Since the geometry has only one U(1) direction fibered over AdS$_2$ base, there should exist only one chiral CFT reproducing the black hole entropy.  Using the metric  (\ref{NHads2}), we get
\be\label{cc}
c=6q_2p\;. 
\ee
which agrees with the result in (\ref{clcr}). Note that the difference in left and right central charges in (\ref{clcr}) is the consequence  of taking into account  the Chern-Simons term.

 Because of the analysis which will be performed in the next section, we assume the bulk entropy scales like $\epsilon$ with $\epsilon \rightarrow 0$. Given the emergence of local AdS$_3$ throats for the EVH black hole, we are interested in matching the AdS$_3$/CFT$_2$  to the Kerr/CFT predictions discussed above.  

The EVH limit, which is given by  (\ref{limitEVH}) and (\ref{chargeEVH}),  is corresponding to the finite central charge ({\ref{cc}), but vanishing level $L_0 -\frac{c}{24}$. 
From the AdS$_3$ perspective,  this corresponds to keeping the gap in the dual CFT finite and sending the level to zero. Notice the CFT temperature scales like the entropy $T\sim S\sim\epsilon$.

The EVH limit sends the level in the CFT to zero. According to the conventional AdS/CFT correspondence, this would
mean that the system is dual to the zero mass BTZ black hole or the pure AdS$_3$, depending on the boundary conditions. 

However this is true, if we do not scale the central charge or the mass gap consequently. It has been shown in \cite{deBoer:2011zt}  that if together with reducing the energy injected in a system we also reduce the mass gap by increasing the central charge, we may remain with a non-trivial dynamical system. 

In the EVH/CFT proposal, in fact we are prescribing to send the mass gap to zero or central charge to infinity too. We shall discuss this issue in section 6. 

\section{Vanishing horizon limits for black holes}
In this section, we will study some characterizations of extremal vanishing horizons black hole solutions among the solutions we reviewed in the previous section.  To get the EVH limit of solutions,  i. e.  $S_{\mathrm{BH} } \sim T_{\mathrm{H}}\sim 0$,  we must take the following limit
\be\label{limitEVH}
\prod \cosh \delta_i = C\epsilon^{-2},\quad m=\mu \epsilon^2,\quad\quad \epsilon\rightarrow 0.
\ee
This can be done in various ways, corresponding to taking a large limit of some or all $\delta_i$ parameters.  If we assume all $\delta_i$ are large and of the same order, from (\ref{BHET}) and (\ref{limitEVH}) we find that all charges and mass are small. The geometry is a small deformation above the product of five-dimensional Minkowski space and five-torus.

  In order to take limit (\ref{limitEVH}), while keeping mass and some charges of black hole finite and non-vanishing, we assume one of $\delta_i$'s, $e.g.$ $\delta_1$ is finite, and two others are large and the same order. This is corresponding to take two charges of the black hole are generic of the same order of magnitude while the other charge is very small.  We take
 \be
 \label{chargeEVH}
\sinh\delta_2 =s_2 \epsilon^{-1}, \quad \sinh\delta_3 =s_3 \epsilon^{-1}  
\ee
   In this limit the black hole entropy and temperature scale to zero with the same rate as $\epsilon$. One of the electric charges, $i.e. \; Q_1$ vanishes while the other electric and magnetic charges and mass of the black hole  remain finite.
   
One may study the near-horizon limit of the geometry obtained in the EVH limit. In order to do that, let us consider the limit (\ref{limitEVH}) and apply the scaling
\bea
r=\frac{2\mu s_2s_3\sqrt{\epsilon}}{R_{\mathrm AdS_3}}  \rho,\quad t=\frac{\tau}{R_{\mathrm AdS_3}\sqrt{\epsilon}},\quad x_5=\frac{\tilde{x}_5}{\sqrt{\epsilon}} ,
\eea
with $\rho, \; \tau$ and $\tilde{x}_5$ held fixed. In this limit, the metric (\ref{10ds}), after transforming into the Einstein frame,  takes the form
\bea
\label{EVH10d}
&& ds^2=-\frac{\rho^2}{R_{\mathrm AdS_3}^2} d\tau^2+\frac{R_{\mathrm AdS_3}^2}{\rho^2}d\rho^2 +\rho^2d\tilde{x}_5^2 +R_{\mathrm AdS_3}^2 d\Omega_3^2+\sqrt{\left|\frac{s_2}{s_3}\right| }\sum_{i=6}^9 dx_i^2 \nonumber  \\ 
&& \label{NEH1} H_{\tau\rho\tilde{x}_5}= -4\mu s_3^2R_{\mathrm AdS_3}^{-3} \rho ,\quad H_{\theta\phi\psi}=4\mu s_3^2\sin\theta\cos\theta, \quad e^{\Phi_{10}}=\frac{s_3^2}{s_2^2},
\eea

where $\tilde{x}_5 \in [0,\sqrt{\epsilon} ] $ and AdS$_3$ radius is given by
\be
R_{\mathrm AdS_3}^4=4\mu^2s_2s_3^3
\ee
As we see the near-horizon metric (\ref{EVH10d}) is exactly of the form that was outlined and discussed in \cite{SheikhJabbaria:2011gc}. In this case, however, the AdS$_3$ radius and the value of the dilaton and gauge fields are determined by the value of the charges $Q_2, Q_3$, defining the EVH black hole \cite{deBoer:2011zt,myworks2}. Although not implied by the equations of motion on the near-horizon geometry, the value of all parameters of the near-horizon configuration are fixed by the charges defining the full EVH black hole, once it is extended out of the horizon and to the asymptotic flat region. In this sense, the EVH black hole shows attractor behavior. We stress that the AdS$_3$ throat in (\ref{EVH10d}) in the near-horizon limit of the EVH black hole is a pinching AdS$_3$, because the circle inside AdS$_3$,  $\tilde{x}_5$ has a vanishing periodicity $\sqrt{\epsilon}$.

One may also study the near-horizon limit of near-EVH black hole. This is corresponding to a small perturbation above two charge black hole by turning on third charge infinitesimally.  To this end, let us consider limit (\ref{limitEVH}) together with the following scaling

\be
\label{rescale2}
r^2=(4\mu^2 s_2^2 s_3^2R_{\mathrm AdS_3}^{-2}\rho^2 -2\mu \sinh^2\delta_1)\epsilon^2,\quad t=\frac{\tau}{R_{\mathrm AdS_3} \epsilon},\quad x_5=\frac{\tilde{x}_5}{\epsilon}
\ee
Taking the limit $\epsilon \rightarrow 0$, we obtain the following geometry

\be
\label{NHBTZ}
ds^2=-\frac{F(\rho)}{R_{\mathrm AdS_3}^2}d\tau^2+\frac{R_{\mathrm AdS_3}^2}{F(\rho)}d\rho^2 +\rho^2 ( d\tilde{x}_5+\frac{\rho_+\rho_-}{R_{\mathrm AdS_3} \rho^2}d\tau )^2 +R_{\mathrm AdS_3}^2 d\Omega_3^2+\sqrt{\left|\frac{s_2}{s_3}\right|}\sum_{i=6}^9dx_i^2
\ee
where
\bea
F(\rho)=\frac{(\rho^2-\rho_+^2)(\rho^2-\rho_-^2)}{\rho^2}
\eea
and $\rho_{\pm}$ are given by
\bea
\rho_+^2=\frac{R_{\mathrm AdS_3}^2  \cosh^2\delta_1\;}{2\mu s_2^2s_3^2},\qquad  \rho_-^2=\frac{R_{\mathrm AdS_3}^2 \sinh^2\delta_1\;}{2\mu  s_2^2s_3^2},
\eea

We note that taking the above near-horizon near-EVH limit does not change the entropy of the original black hole. To see the latter, one may reduce the 10d gravity theory (\ref{action10dE}) to three dimensions. The 3d Newton constant is computed in the standard fashion, 
\bea
\frac{1}{16\pi G_{10}} \int d^{10}x\sqrt{-{\mathrm det} \; g_{10}}\;( R_{10} + \cdots)=\frac{1}{16\pi G_{3}} \int d^{3}x\sqrt{-{\mathrm det}\; g_{3}}\;( R_{3} + \cdots)
\eea
where {\it dots} denotes the action for the other fields.  Proceeding in this way, one finds
\bea
\frac{1}{G_3} = \frac{8\pi^2 \mu^2 s_2^2 s_3^2R_{\mathrm AdS_3}^{-1}}{ G_{10}} 
\eea

The Bekenstein-Hawking entropy of the pinching BTZ solution to this 3d theory is then given by
\bea
S_{3d}=\frac{\epsilon \rho_+}{4G_3}=\frac{\epsilon R_{\mathrm AdS_3} \cosh \delta_1}{G_{10}}
\eea
which is identical with the entropy of the original near-EVH black hole, given by charges (\ref{chargeEVH}).  It is also instructive to compare the Hawking temperatures of the original near-EVH  black hole and that of the pinching BTZ:
\bea
T_{{\mathrm BH}}=\frac{\epsilon\; {\mathrm sech} \;\delta_1 }{2\pi  \sqrt{2\mu} \; s_2s_3} =R_{\mathrm AdS_3} \epsilon T_{\mathrm BTZ}
\eea
where $T_{\mathrm BTZ}= \frac{\rho_+^2-\rho_-^2}{2\pi\rho_+}$. The prefactor $\epsilon$ is expected from rescaling (\ref{rescale2}).

In \cite{Prester:2009mc},  it has been shown for several examples of string backgrounds containing AdS$_3$ factor, how to calculate conformal central charges for BPS and non-BPS solutions using the complete tree-level effective action by taking into account all higher-derivative terms. In particular the AdS$_3 \times$S$^3$ solution in heterotic string theory has been studied there.  

It has been argued that after adding appropriate corrections terms, including the Chern-Simons into the action  (\ref{action10dE}), the equations of motion  admit an AdS$_3 \times$S$^3\times$T$^4$ geometry. Using this solution one can read off central charges which are given by
\be
c_L+c_R=12|q_2 p|+18|q_2|,\quad c_L-c_R=18|q_2|.
\ee
%

\section{EVH/CFT description}
In Section 4, we have studied near-horizon limit of near-EVH black holes and shown that we generically obtain an AdS$_3$ throat. We also have seen that the entropy of the original near-EVH black hole is parametrically equal to the entropy of the BTZ geometry obtained in the near-horizon limit. 

The appearance of the AdS$_3$ throat in the near-horizon of the EVH black holes is very suggestive of the existence of a 2d-CFT dual to physics on this geometries. Then it is standard to reproduce the gravitational entropy of an AdS$_3$ throat by using Cardy formula. However we should note that, what we obtain in the near-horizon limit is not a round AdS$_3$, it is a {\it pinching orbifold} of AdS$_3$. 

In the presence of a non-trivial pinching, using results from the dual CFT may be a bit more subtle. When we are computing the mass and angular momentum of the 3d pinching geometry (\ref{NHBTZ}), the non-trivial periodicity of the S$^1$ circle should take into account. Using the same argument, the Brown-Henneaux central charge will also acquire a linear $\epsilon$ dependence
\bea
c_{\mathrm AdS_3}=\frac{3R_{\mathrm AdS_3}}{2G_3}\epsilon=6q_2p \epsilon,
\eea

which agrees with the space-time CFT approach discussed in \cite{Martinec:2001cf} .  The near-horizon limit (\ref{rescale2}) can be assumed correspond to an IR limit of a 2d-CFT, in which both chiral sectors are decoupled and we are left with no dynamics  \cite{EVH-BTZ} .  This is the consequence of the pinching appearing in the near-horizon geometry (\ref{NHBTZ}). So, we need to have proposals for resolving the `pinching orbifold'.

Note that the above conclusion is true while holding NewtonÕs constant fixed.  However if we allow NewtonÕs constant to scale to zero as the near-horizon limit is taken, one can keep the central charge and the entropy finite  \cite{EVH-BTZ}.  In  \cite{SheikhJabbaria:2011gc} , it has been argued that, the pinching can be removed by scaling the Newton coupling constant to zero. In addition, as long as the seven dimensional part of near-horizon geometry remains finite, both 3d and 10d Newton constants scale to zero in the same way. So we accompany the already double scaling near-EVH near-horizon limit of the previous sections 
 \bea
 A_h, T, G_N \rightarrow 0, \quad \frac{A}{G_N}, \frac{A_h}{T}\;\; finite,
 \eea
where $A_h/G_N$ is the entropy of the EVH black hole and the ratio $A_h/T$ is proportional to the central charge of the dual CFT.  The dual 2d-CFT, after resolution of the pinching orbifold singularity, has a finite central charge $c$, which is given by
\bea
c_{CFT}=6q_2p,
\eea

The identification of $L_0$ and $\bar{L}_0$ in terms of the BTZ parameters can be done in the standard way, i.e.
\bea
L_0-\frac{c}{24}=\frac{(\rho_+-\rho_-)^2}{16G_3},\quad \bar{L}_0-\frac{c}{24}=\frac{(\rho_++\rho_-)^2}{16G_3}
\eea

Cardy formula is consistent with Bekenstein-Hawking, as usual.
\bea
S_{\mathrm Cardy}=2\pi \sqrt{\frac{c}{6}(L_0-\frac{c}{24}) }+2\pi \sqrt{\frac{c}{6}(\bar{L}_0-\frac{c}{24}) } 
\eea

Then  BTZ black hole can be considered a thermal state in the 2d-CFT, which specified above at temperature $T_{\mathrm BTZ} = \frac{\rho^2_+-\rho_-^2}{2\pi\rho_+} $. With this identification and recalling our
earlier discussions, it is then obvious that Cardy formula which produces the BTZ black hole entropy, correctly reproduces the near-EVH black hole entropy as well.

The connection between left and right sectors of our EVH/CFT's and the chiral CFT of Kerr/CFT  lies within the Discrete Light Cone Quantization (DLCQ )proposal \cite{SheikhJabbaria:2011gc}, at least for near-EVH extremal black holes. 

Any extremal near-EVH black hole in the near-horizon limit goes to an extremal BTZ black hole throat, and this BTZ may be thought as excitation of an AdS$_3$ throat of an EVH black hole. 
Then the Kerr/CFT proposal, at least for near-EVH black holes, can be derived from the EVH/CFT proposal, if we take yet another near-horizon limit over the extremal BTZ.  

In \cite{us} it has been shown that taking near-horizon limit over an extremal BTZ black hole corresponds to  DLCQ of the dual 2d-CFT. In the DLCQ of a given 2d-CFT we basically freeze out  a chiral sector of the CFT and we remain with a "chiral CFT".  Our proposal is that the chiral CFT of Kerr/CFT type duality should be viewed as a DLCQ of a non-chiral generic 2d-CFT                                                            .
\section{Conclusion}

In this paper we have analyzed the near-horizon geometry of the five-dimensional static charged black hole solution in heterotic string theory compactification. The solution is parametrized by mass, two electric charges and one magnetic charge. Besides the three U(1) gauge fields, the background includes two scalar fields which represent components of the string metric and dilaton along a circle of the internal five-dimensional torus. 

By analyzing the near-horizon field configuration at the vanishing limit of the horizon's area and Hawking temperature we have found that the near-horizon geometry develops an AdS$_{3}$ throat. The same approach has been used in \cite{SheikhJabbaria:2011gc} to study a general rotating black hole in Einstein gravity coupled to one gauge field and one scalar field. The AdS$_3$ throat which appears in the near-horizon limit is a pinching AdS$_3$ = AdS$_3$/Z$_K$, $K \rightarrow \infty$. 

Furthermore, we have shown that the near-horizon limit of the near-EVH black holes has a pinching BTZ factor. The appearance of the AdS$_3$ factor in the near-horizon geometry is a good indication for trying to establish the EVH/CFT. To resolve the pinching issue, following the proposal in \cite{SheikhJabbaria:2011gc}  we accompany the near-EVH near-horizon limit by a particular $G_N\rightarrow 0$ limit.

Although it is true that we are scaling the 10d Newton constant to zero, we do have a 3d gravity theory with finite Newton constant. It is somehow like taking the $\alpha' \rightarrow 0$ limit in string theory or near-horizon limit in AdS/CFT. This limit is needed to throw out some of the degrees of freedom while keeping some others. This will simplify the theory and reduces the theory to something simple, tractable, but still non-trivial.

Explicitly, we proposed the following triple scaling limit: horizon area, Hawking temperature and $G_N\rightarrow 0$, keeping the ratios horizon area to temperature and horizon area to Newton coupling constant fixed. It implies a particular duality between 2d-CFTs and its orbifold: 2d-CFT  with central charge $c$ on cylinder $R\times S^1$ is dual to 2d-CFT  with central charge $cK$ on$R\times S^1/Z_K$ in the large K limit. 

Let us also comment on the connection between the 2d-CFT  description we discussed in the previous sections and the Kerr/CFT proposal. A possible connection between the two can come along the lines of \cite{us}, \cite{EVH-BTZ} and discussed with some details in  \cite{SheikhJabbaria:2011gc}: The EVH/CFT in the DLCQ description reproduces Kerr/CFT. For the above to work one should, however, extend the validity of our EVH/CFT proposal beyond the strict near-EVH region. In other words, generic extremal black holes may be viewed as excitations above the EVH black hole, when one sector of the dual 2d-CFT  has been excited.

The Kerr/CFT central charge $c$ is finite before scaling Newton constant. In the regime of charges which is considered in the near-EVH limit, there is an AdS$_3$ throat emerging in the near-horizon gravitational description.  This allows us to identify the CFT as the parent 2d-CFT  discussed in Section 6.  The pinching orbifold is dual to a thermal state in this parent CFT exploring very low energies where $L_0-\frac{c}{24}\rightarrow 0$ at low Frolov-Thorne temperatures $T\rightarrow 0$.  In \cite{Martinec:2001cf}, authors  discussed the space-time conformal algebra perspective which is corresponding to the Ôlong string sectorÕ of this parent CFT.  This picture is the same in sprit as the one discussed in  \cite{SheikhJabbaria:2011gc}  for a similar situation in a much simpler setting of EVH BTZ black hole.  In this example, by taking the near-horizon limit  then going to near-EVH region, without scale the central charge, one gets the null self-dual orbifold. However, if we take the near-EVH limit first, then going to the near-horizon limit we end up with a pinching AdS$_3$ orbifold.  One may start with a parent CFT with finite central charge and exploring low-energy excitations. At the $G_N\rightarrow 0$ limit, one can keep some non-trivial dynamics by considering the double scaling limit, i.e. $L_0-\frac{c}{24} \rightarrow 0$ and $c \rightarrow \infty$, keeping the entropy finite.

The EVH black holes are not limited to static ones. In \cite{myworks} we have studied some examples of stationary EVH black holes. Within the class of heterotic black hole solutions we have a more general family of EVH black holes which involve rotation as well as electric and magnetic charges. This class of charged-rotating EVH black holes and rotating black holes in heterotic string theory compactification will be studied in a future publication \cite{future}.

\section*{Acknowledgements}
I would like to thank Sameer Murthy, Shahin Sheikh-Jabbari and Joan Sim\'on for useful comments and discussion.  This work was supported by the National Research Foundation of Korea Grant funded by the Korean Government (NRF-2011- 0023230).

\end{document}